\documentclass{article}

\usepackage{arxiv}

\usepackage[utf8]{inputenc} % allow utf-8 input
\usepackage[T1]{fontenc}    % use 8-bit T1 fonts
\usepackage{hyperref}       % hyperlinks
\usepackage{url}            % simple URL typesetting
\usepackage{booktabs}       % professional-quality tables
\usepackage{amsmath,amsfonts,amssymb}       % blackboard math symbols
\usepackage{nicefrac}       % compact symbols for 1/2, etc.
\usepackage{microtype}      % microtypography
\usepackage{graphicx}
\usepackage[square,numbers]{natbib}
\usepackage{doi}

\title{Large FOV short-wave infrared meta-lens for scanning fiber endoscopy}

\author{Ningzhi Xie \\
	Department of Electrical and Computer Engineering\\
	University of Washington, Seattle, WA 98195 \\
	\texttt{nzxie@uw.edu} \\
	\And
	Matthew D. Carson\\
	Department of Mechanical Engineering\\
	University of Washington, Seattle, WA 98195 \\
	\texttt{mdc34@uw.edu} \\
	  \AND
	Johannes E. Fr{\"o}ch \\
	Department of Electrical and Computer Engineering\\
        Department of Physics \\
	University of Washington, Seattle, WA 98195 \\
	\texttt{jfroech@uw.edu } \\
	  \And
	Arka Majumdar \\
	Department of Electrical and Computer Engineering\\
        Department of Physics \\
	University of Washington, Seattle, WA 98195 \\
	\texttt{arka@uw.edu} \\
	\And
	Eric Seibel\\
	Department of Mechanical Engineering\\
	University of Washington, Seattle, WA 98195 \\
	\texttt{eseibel@uw.edu} \\
        \And
	Karl F. B\"ohringer \\
	Department of Electrical and Computer Engineering\\
        Department of Bioengineering \\
	University of Washington, Seattle, WA 98195 \\
	\texttt{karlb@uw.edu} \\  
}

\date{}

\hypersetup{
pdftitle={Large FOV short-wave infrared meta-lens for scanning fiber endoscopy},
pdfsubject={optics},
pdfauthor={Ningzhi Xie},
pdfkeywords={endoscope, scanning fiber, meta-lens, large FOV, ultra-small optics,infrared angioscope},
}

\begin{document}
\maketitle

\begin{abstract}
The scanning fiber endoscope (SFE), an ultra-small optical imaging device with a large field-of-view (FOV) for having a clear forward view into the interior of blood vessels, has great potential in the cardio-vascular disease diagnosis and surgery assistance, which is one of the key applications for short-wave infrared (SWIR) biomedical imaging. The state-of-the-art SFE system uses a miniaturized refractive spherical lens doublet for beam projection. A meta-lens is a promising alternative which can be made much thinner and has fewer off-axis aberrations than its refractive counterpart. We report an SFE system with meta-lens
working at 1310nm to achieve a resolution ($\sim 140\mu m$ at the center of field and the imaging distance of $15mm$), FOV ($\sim 70 \circ$), and depth-of-focus (DOF $\sim 15mm$), which are comparable to a state-of-the-art refractive lens SFE. The use of the meta-lens reduces the length of the optical track from $1.2mm$ to $0.86mm$. The resolution of our meta-lens based SFE drops by less than a factor of $2$ at the edge of the FOV, while the refractive lens counterpart has a $\sim 3$ times resolution degradation. These results show the promise of integrating a meta-lens into an endoscope for device minimization and optical performance 
 improvement.
\end{abstract}

% keywords can be removed

\section{Introduction}
\label{sect:intro} 
In diagnosis of diseases and surgery in the cardio-vascular system, it is essential to have a clear view into the interior of blood vessels. An endoscope with an extremely small diameter $(\lesssim 1 mm)$ and short rigid tip length $(\lesssim 10 mm)$ can provide a viable solution. To that end, scanning fiber endoscopes (SFEs) have already been demonstrated as one of the smallest forward viewing endoscopes due to their single point scanning based optical sensing mechanism \cite{SFE1,SFE2}. Alternative techniques include imaging through a lens on a coherent fiber bundle (CFB) array \cite{FB1,FB2} or a camera sensor chip-on-tip \cite{COT1}, which are difficult to miniaturize while maintaining high performance of large field of view (FOV) and high resolution. SFEs also feature a large depth-of-focus (DOF) ($\sim 10mm$), have a larger FOV and more number of pixels than CFB arrays and camera sensors chip-on-tip \cite{Endoscope_Rv}. 

The short wavelength infrared (SWIR) wavelength regime is of particular interest for the surgery-assisting endoscopy used inside blood vessels, as they are the shortest wavelength ranges to ensure both reduced scattering and low water absorption. This enables an infrared endoscope design \cite{IR_endocope} to see through blood without requiring any additional clearing. In fact, an SWIR SFE at $1310nm$ has already been reported \cite{SFE3}: here a compound refractive spherical lens is used to project the beam emitted from a scanning single mode fiber to a large range of angles ($-35^{\circ}$ to $35^{\circ}$ relative to the optical axis). The size of this optical element is one of the major limitation factors for further miniaturization of the rigid tip length of the SFE system. Additionally, the spherical lens also suffers from large aberration, when the beam is projected to $>25^{\circ}$ off-axis angles. 

\begin{figure}
\begin{center}
\begin{tabular}{c}
\includegraphics[width=16cm]{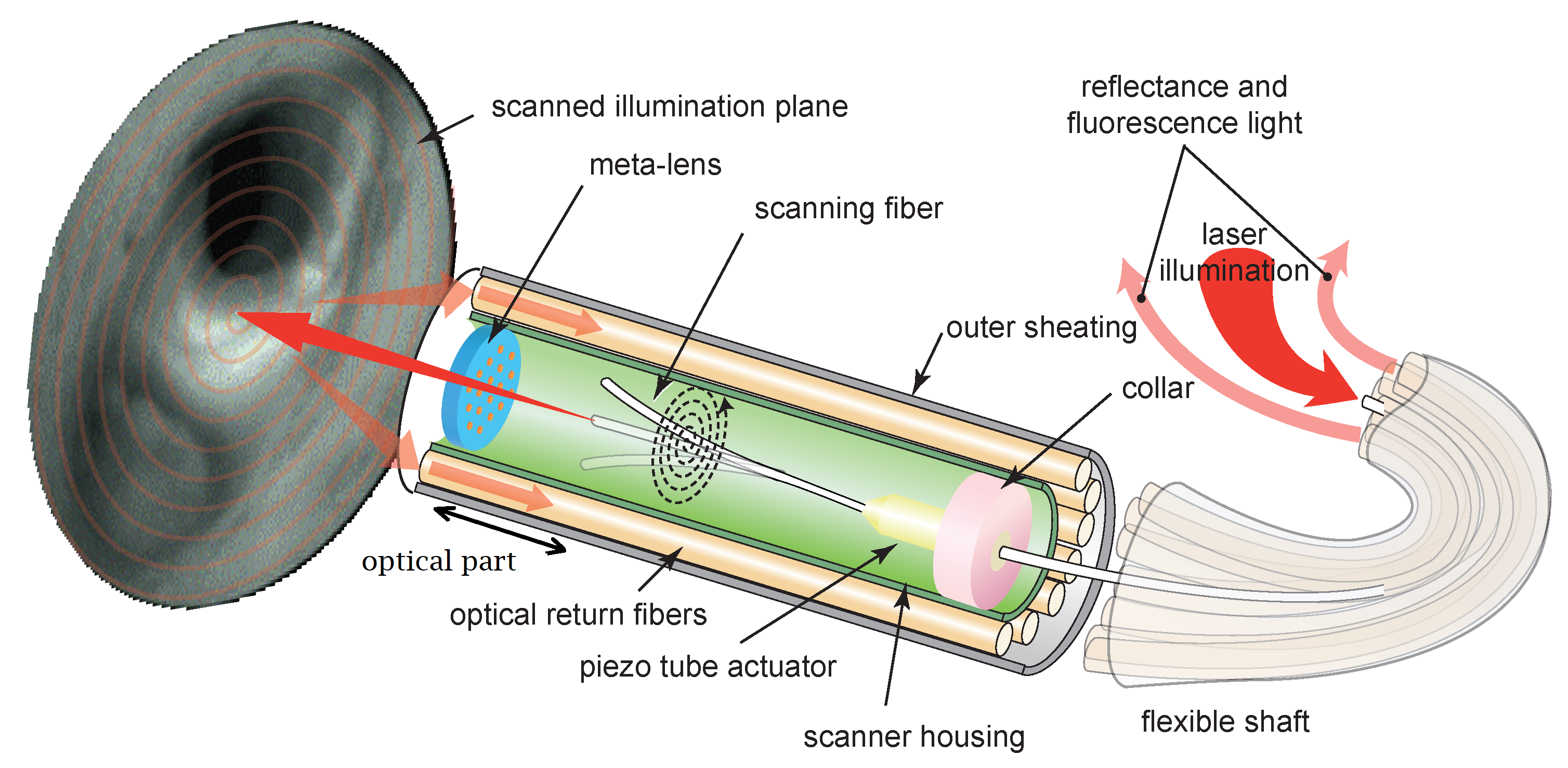}
\end{tabular}
\end{center}
\caption 
{\label{Fig_SFE_sketch} \textbf{The schematic of a meta-lens based SFE system.}
A suspended single mode fiber is driven by a piezo tube to follow a spiral
scanning path, and the beam emitted from the fiber tip at different locations
is projected to different angles by a meta-lens to illuminate the object. As the fiber scans following a spiral path, the beam scans across the whole FOV. The object diffusively reflects the scanning beam, which is collected by the return fiber to form an image of the object. } 
\end{figure}

Meta-optics has recently emerged as a promising alternative to drastically miniaturize optical elements \cite{MS1}. Meta-optics consist of quasi-periodic arrays of sub-wavelength scatterers, each of which can be designed to impart desired phase, amplitude, polarization and spectral control of light. The ability to engineer any phase-mask with sub-wavelength resolution makes meta-optics an excellent candidate to create ultra-thin freeform optics \cite{MS2}. Meta-optical lenses, commonly known as meta-lenses, are extremely thin (thickness approaching $\sim \mu m$ for SWIR), and can be fabricated to have very small area. By intelligently designing the phase-mask, meta-optics can correct for aberrations at large off-axis angles, without using multiple optical components \cite{MS3}. This ability of meta-optics has already inspired several works on meta-optical endoscopy \cite{MSE1,MSE2,MSE3}. Here, we report an SFE system with a meta-lens working at $1310nm$ to achieve comparable resolution, FOV, and DOF as the state-of-the-art, while minimizing the rigid tip-length and reducing the aberration at large off-axis angles. Fig.~\ref{Fig_SFE_sketch} shows the schematic of our SFE system. This imaging system has a full FOV of $70^{\circ}$ and a DOF of $\sim 15mm$ (in this range the resolution drops less than a factor of 2). The replacement of the refractive compound lens with a single meta-lens reduces the length of the optical track of the SFE from $1.2mm$ to $0.86mm$. Further reduction to $~0.4mm$ is possible with a thinner substrate. The spatial resolution at the center of the field is $140\mu m$ at an illumination distance of $15mm$, which corresponds to an angular resolution of $0.53^{\circ}$. This resolution drops by less than a factor of 2 at the edge of the FOV.

\section{Design and fabrication of the meta-lens}
\subsection{Design of the meta-lens}

The purpose of the meta-lens in our SFE system is to project the beam emitted from the tip of the scanning fiber to the illumination plane. Specifically, the beam incident at different positions of the meta-lens is guided to different angles. The SFE should have a wide range of $\sim 20mm$ imaging distance, wherein the projected beam maintains a small diameter. Therefore, the projected beam should be collimated. This functionality can be realized by a single layer optical component, which can be approximated by a hyperboloid focusing lens. This lens images the fiber tip at its back focal plane to infinitely far. Therefore, we assume the initial phase profile of the meta-lens as:

\begin{equation}\label{phase_profile_1}
\begin{split}
\Phi_0 =  \frac{f - \sqrt{\rho^2+f^2}}{\lambda} \cdot 2 \pi
\approx - \frac{\pi}{f \lambda} \rho^2 
\approx - \frac{\pi}{d \lambda} \rho^2
\end{split}
\end{equation}
where $\rho$ is the radius from the center of the lens, $\lambda$ is the wavelength ($1310nm$), $f$ is the focal length of the lens, $d$ is the distance between the fiber tip and the meta-lens, $f \approx d$.

The phase profile of the meta-lens is then optimized using the ray-tracing simulator \emph{Zemax Optical Studio}. The meta-lens is represented by a binary2 phase mask whose phase profile is expressed as:

\begin{equation}\label{phase_profile_b2}
\Phi = \sum_{k=1}^n A_k \rho^{2k}
\end{equation}

where $k$ is the number of order, and $A_k$ are the corresponding coefficients that we optimize. To ensure centrosymmetry, only even order exponents are considered. We set $n=3$, as the first 3 terms turn out to be sufficient for our optimization. We use $\Phi_0$ (i.e., $A_1=-\frac{\pi}{d\lambda},A_2=A_3=0$) as the starting point for optimization.

\begin{figure}
\begin{center}
\begin{tabular}{c}
\includegraphics[width=16cm]{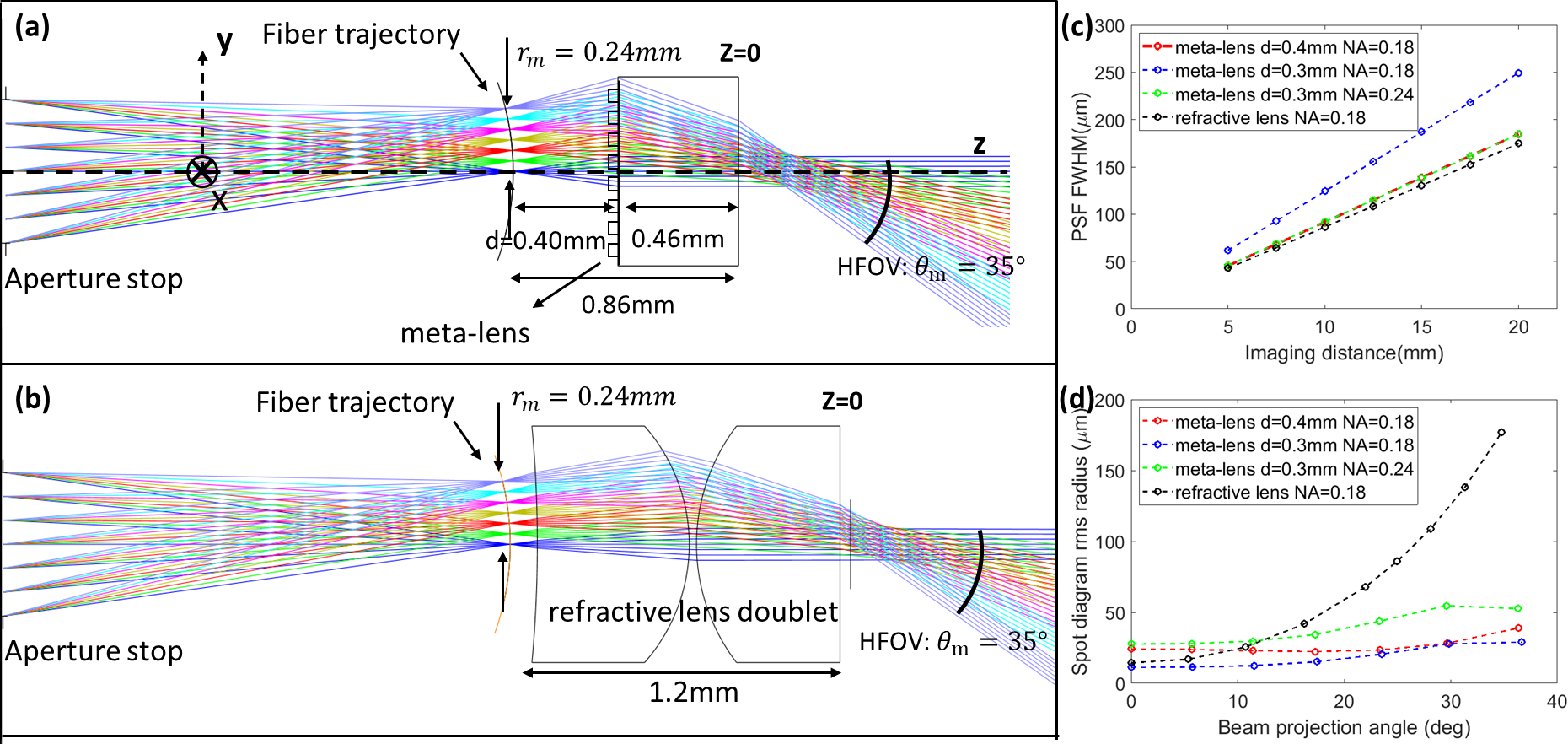}
\end{tabular}
\end{center}
\caption 
{ \label{Fig_raytracing} \textbf{(a)} Cross-section of the scanning fiber imaging system optimized in Zemax using ray-tracing. The object plane simulates the trajectory of the fiber tip during scanning. Each object point position represents one fiber tip position. 7 object points are placed along the $y$ axis with various $y$ values from $0$ to $r_m = 0.24mm$ with an increment of $0.04mm$. An aperture stop is placed ahead of the object surface, which reflects the limited emission solid angle of the beam emitted from the fiber $(\mathit{NA} = 0.18)$. The aperture stop size, the object plane curvature, and the distance between the stop and the object are chosen to simulate the scanning behavior of the fiber. A binary2 phase mask is used to model the meta-lens, which is on a sapphire substrate with the refractive index $n = 1.75$ at the operation wavelength $\lambda = 1310nm$. Rays in different colors represent the beam emitted from the fiber tip at different positions and be projected to different angles. \textbf{(b)} The design of the state-of-the-art scanning fiber imaging system with the same fiber scanner, but using a refractive spherical lens doublet, which has the same FOV of $35^{\circ}$, but a longer length of $1.2mm$ compared to $0.86mm$ in (a). \textbf{(c)} FWHM of the PSF calculated by Huygens method as a function of the imaging distance, at the center of the field ($x=0,y=0$). \textbf{(d)} Spot diagram rms radius as a function of the beam projection angle for different SFE designs using meta-lens or refractive lens doublets, at the imaging distance $z=20mm$. } 
\end{figure}

Fig.\ref{Fig_raytracing}a shows the cross-section of our SFE system in the ray-tracing simulation. The trajectory of the fiber tip during scanning is represented by the object plane, whose geometry is predicted by a theoretical model\cite{FB_scanner}. The distance between the fiber and lens $d$, the radius of the object plane $r_m$ (which represents the range of motion of the fiber), and the numerical aperture $\mathit{(NA)}$ of the aperture (which is the $\mathit{NA}$ of the fiber) are important design parameters that determine the FOV, DOF, and resolution of the SFE system. $d$ and $r_m$ determine the half-FOV $\theta_m$ by $\theta_m \sim \arctan(r_m/d)$. A smaller $d$ is desired not only because it reduces the length of the optical track of the SFE system, but also requires a smaller fiber scanning range $r_m$ to achieve a certain half-FOV. However, the width of the beam $w$ on the meta-lens ($w \sim d \cdot \mathit{NA}$) also decreases with $d$, leading to a larger diverging angle ($\sim \frac{\lambda}{\pi w} $) of the projected beam, which degrades the DOF and resolution of the SFE system. The fiber we use has an $\mathit{NA}$ of $0.18$. We set $d=0.4mm$ to ensure a relatively short optical track length and acceptable projected beam diverging angle, and $r_m = 0.24mm$ to get a half FOV of $35^{\circ}$. To optimize the SFE performance, the coefficients $A_1,A_2,A_3$ are adjusted to minimize the spread of the ray at $z = 20mm$. Specifically, we choose $7$ uniformly spaced points along the y-axis in the object plane, and minimize the average radius of the resulting $7$ spots in the image plane. The point spread function (PSF) is the intensity distribution of the beam projected by the meta-lens, which determines the resolution of the SFE system. We used the Huygens method in Zemax to calculate the PSF. 

Compared to a state-of-the-art SFE system with a refractive spherical lens doublet (Fig.\ref{Fig_raytracing}b), our SFE system with the meta-lens shows the same FOV of $70^{\circ}$, a comparable full-width-at-half-maximum (FWHM) of the PSF at the center of the field (Fig.\ref{Fig_raytracing}c), and a much smaller rms radius of the spot diagram at the projection angle $\theta > 20^{\circ}$ (Fig.\ref{Fig_raytracing}d). This means that our meta-lens can reduce the spherical aberration at large projection angles. The use of the meta-lens also reduces the optical track length in the SFE from $1.2mm$ to $0.86mm$. This length can be further reduced to $\sim 0.4mm$ by using a thinner substrate ($\sim 0.1mm$) and a smaller $d \sim 0.3mm$ between the fiber and the lens. As shown in Fig.\ref{Fig_raytracing}c, reducing $d$  increases the beam diverging angle, which indicates a degraded resolution and DOF. This can be mitigated by increasing the $\mathit{NA}$, at the cost of a larger aberration at large beam projection angles. 

\subsection{Meta-atom pattern design}
\begin{figure}
  \begin{center}
  \includegraphics[width=16cm]{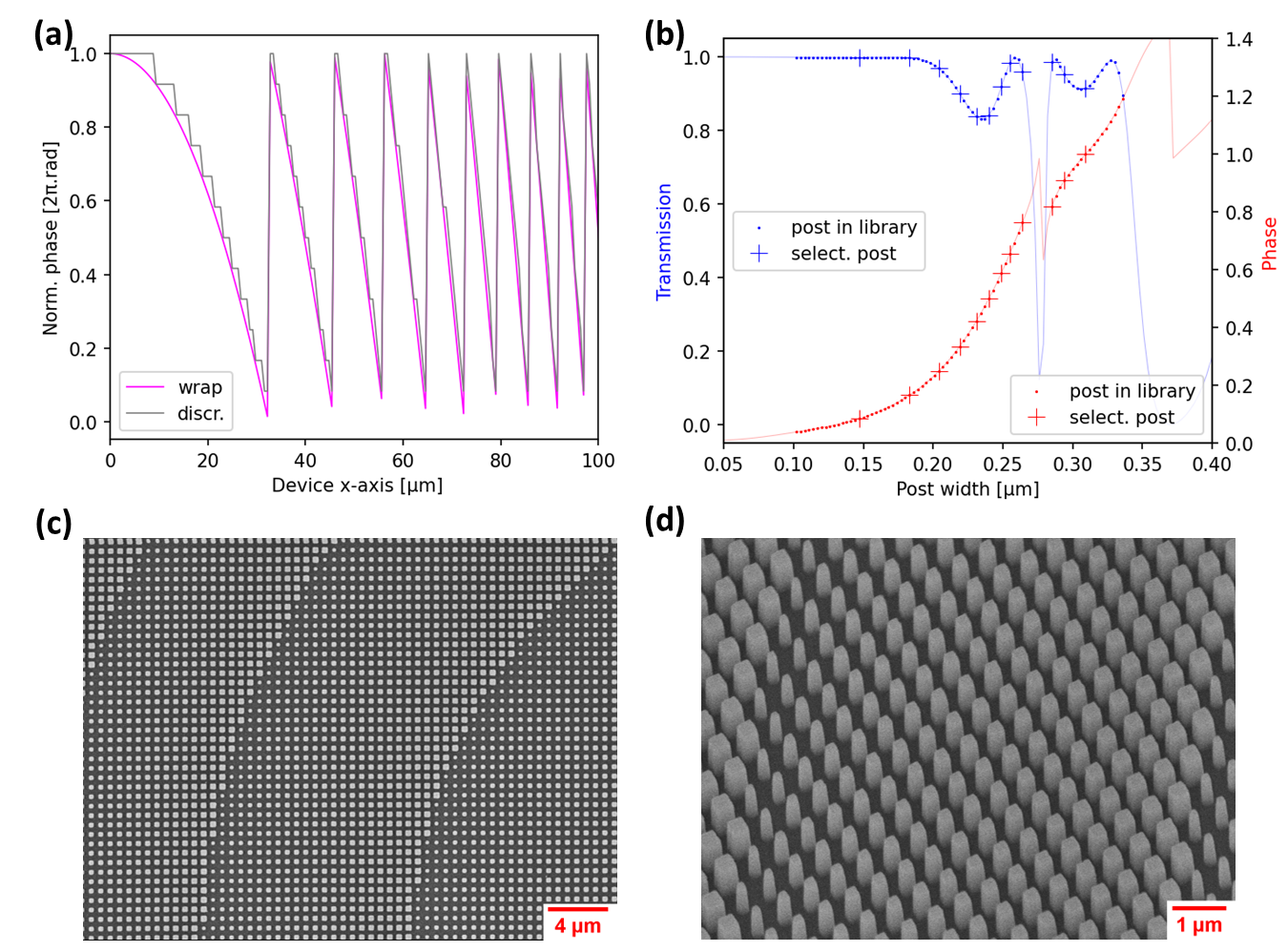}\\
 \caption{\label{Fig_meta-lens}\textbf{(a)} The wrapped and discretized (12 equally spaced levels) phase profile cross-section ($0<x<100\mu m$) of the meta-lens designed in Fig.~\ref{Fig_raytracing}a. \textbf{(b)} Nanopost phase and transmission response calculated by rigorous coupled wave analysis (RCWA), with cSi nanoposts on $460\mu m$ thick Al$_2$O$_3$ substrate at a wavelength of $1310nm$. The height of the nanoposts are $h=1000nm$ and the periodicity of the nanopost array is $p=600nm$. \textbf{(c)} Scanning electron microscope (SEM) images of the fabricated meta-lens from the top. \textbf{(d)} SEM image at the oblique angle of $40^{\circ}$.}
  \end{center}
\end{figure}

The phase profile of the meta-lens obtained from the ray-tracing simulation is wrapped and further discretized into 12 levels, as shown in Fig.\ref{Fig_meta-lens}a. These phase levels are realized by a crystalline silicon (cSi) on sapphire (Al$_2$O$_3$) meta-optics, in which the sub-wavelength nanoposts with different sizes act as local scatterer to impose the desired phase response on the incident light. We use rigorous coupled wave analysis (RCWA)\cite{RCWA} to theoretically calculate the phase and transmission response vs the scatterer size, as shown in Fig.\ref{Fig_meta-lens}b. This allows us to pick the right scatterer size at specific locations and thus construct the meta-lens. 

\subsection{Meta-lens fabrication}
The meta-lens is fabricated using an electron beam lithography (EBL) process. A resist (ZEP-520A) on a $1\mu m$-thick crystalline Si-on-sapphire substrate is patterned by EBL, then a $\sim 50nm$ thick Al$_2$O$_3$ hard mask is created by electron beam assisted evaporation followed by resist ($\sim 200$nm thick) lift-off in N-methyl-2-pyrrolidone (NMP) at $90^\circ$C overnight. Subsequently, the crystalline-Si layer is etched by a  fluorine based reactive ionprocess. The scanning elecron microscope (SEM) images of the fabricated meta-lens are shown in Figs.~\ref{Fig_meta-lens}c and ~\ref{Fig_meta-lens}d. 

\section{Characterization of the meta-lens}

\subsection{Experimental setup}

The DOF and resolution of our SFE system are determined by the longitudinal (in the x-z plane) and transverse (in the x-y plane, also the illumination plane) intensity distribution of the beam projected by the meta-lens. Fig.~\ref{Fig_non-scan_measure}(a) shows the schematic of the setup for measuring these beam intensity distributions. A customized fiber scanner is built \cite{FB_scanner,SFE3}. In our SFE, the meta-lens projects the beam emitted from the tip of the scanning fiber to different angles within a total FOV of $70^{\circ}$. A movable IR microscope consisting of an objective  (Mitutoyo Plan Apo NIR 10x, $f=20mm$,$NA=0.26$), a tube lens (Thorlabs AC254-075-B-ML, $f=75mm$), and an IR sensor (WiDy SenS S320 V-ST) is placed in front of the meta-lens along the optical axis to characterize the beam on the focal plane of the microscope. This microscope is mounted on a 3D translation stage, which can be moved either along the optical axis (in z direction) or perpendicular to it (in the x-y plane). By moving the microscope along the z-direction, the beam transverse intensity distribution at different illumination distances $z$ can be measured to construct the longitudinal distribution. By moving the microscope in the x-y plane, the beam intensity distribution at different projection angles can be measured. At beam projection angles $>14^{\circ}$ relative to the optical axis, the objective does not have a large enough NA to capture the light. Therefore, a frosted film is put at the focal plane of the microscope to scatter the beam, and the scattered beam is collected with the microscope.

\subsection{Longitudinal intensity distribution of the beam}

\begin{figure}
  \begin{center}
  \includegraphics[width=16cm]{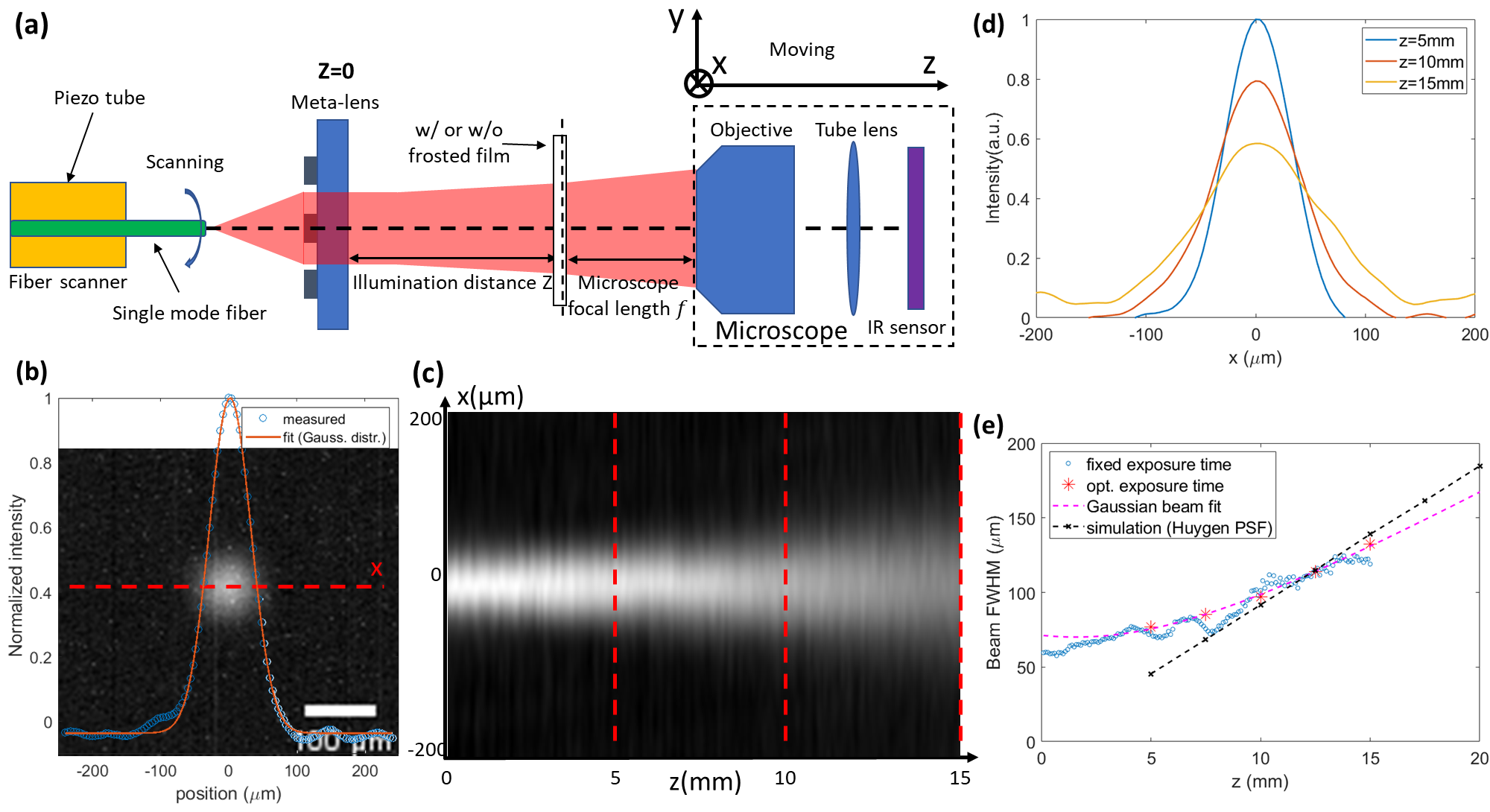}\\
 \caption{\textbf{(a)} Schematic of the measurement setup for meta-lens characterization. \textbf{(b)} Direct image of the beam without fiber scanning at the illumination distance $z=5mm$.  A Gaussian blur (radius = $12\mu m$) is applied to the image to reduce the noise. After that, the intensity distribution along x-axis (the red dashed line) is subtracted and fitted with a Gaussian distribution function. The measured and fitted intensity distribution is plotted on the image.  \textbf{(c)} Longitudinal beam intensity distribution, constructed from a series of images similar to (b) taken at $z$ ranging from 0 to 15mm with an increment of 0.1mm. The beam power from the fiber and the exposure time are fixed. \textbf{(d)} Beam intensity distribution along x-axis for various illumination distance $z$, corresponding to the three red dashed line in (c). \textbf{(e)} FWHM of the beam intensity along x-axis vs $z$ at the center of the field. Blue circles correspond to a series of images with a fixed exposure condition. Red stars correspond to images with dynamically optimized exposure conditions to maximize the signal-to-noise ratio. The magenta dashed line is a fit to the red stars, assuming the beam is a perfect Gaussian beam, according to equation \ref{Gaussian beam fit}. Fitted Gaussian beam waist width is $w_0 = 59.5 \mu m$ (corresponding $\mathit{FWHM} = 70.1\mu m$), while the waist position is at $z=1.58mm$. The black dashed line is the PSF calculated in Zemax according to Huygens method.}\label{Fig_non-scan_measure}
  \end{center}
\end{figure}

We first construct the longitudinal intensity distribution (in the x-z plane) of the beam projected by the meta-lens at $0^{\circ}$ projection angle (i.e., neutral position of the fiber) by translating the microscope along the optical axis and take images of the beam at different $z$. Fig.~\ref{Fig_non-scan_measure}b shows the image of the beam, taken at $z=5mm$. The 1D intensity distribution along the x-axis (the red dashed line) is plotted across the image, and further fitted by a Gaussian distribution function. From further measurements of the 1D distributions along the x-axis at different planes in the range $z=0 - 15mm$, we construct the longitudinal intensity distribution (in the x-z plane), shown in Fig.~\ref{Fig_non-scan_measure}c. Importantly, this directly indicates that the beam diverges only slightly over a long distance of $15mm$. Further 1D intensity distributions for different $z$-distances are plotted in Fig.~\ref{Fig_non-scan_measure}d, showing how the beam shape is maintained over $z = 5mm - 15mm$ while slight broadening. To further characterize this system we use an analytical model based on Gaussian beam propagation, summarized in Fig.~\ref{Fig_non-scan_measure}e, where the FWHM of the 1D intensity distribution as a function of $z$ can be well fitted by using :

\begin{equation}\label{Gaussian beam fit}
\begin{split}
\mathit{FWHM}(z) = w(z) \sqrt{2\times ln2}  \\
w(z) = w_0 \sqrt{1+[(z-z_0)/z_R]^2} \\
z_R = \pi w_0^2 / \lambda
\end{split}
\end{equation}

By fitting, we estimate the beam waist at position $z_0=1.58mm$ with waist width $w_0=59.5\mu m$, which corresponds to a FWHM of $70.1\mu m$, whereas further along the optical axis at $z=15mm$ the beam enlarges to  $FWHM = 135\mu m$. The FWHM of the 1D intensity distribution of the projected beam increases by less than a factor of 2 when $z$ increases from $0$ to $15mm$. Therefore, we define the DOF of our SFE system to be $15mm$. We demonstrate that the beam projected by the meta-lens has a longitudinal intensity distribution close to a Gaussian beam whose beam waist is at the position of the meta-lens, which yields the minimal diverging angle for a certain beam width. This ensures a large DOF of $15mm$ in our SFE system. The measured beam longitudinal distribution slightly deviates from the simulation results calculated in Zemax ray tracing simulation using Huygens method (black dashed line in Fig.\ref{Fig_non-scan_measure}e). We attribute this to the fact that the actual beam from the fiber deviates from our model of a point source.

\subsection{Scanning beam FWHM at various projection angles in the illumination plane}

We characterize the spatial resolution of the SFE system at various angles in the illumination plane by measuring the FWHM of the 1D transverse intensity distribution of the scanning beam. This 1D distribution of the beam can be measured by taking the image of a stable trajectory of the scanning beam with an exposure time exceeding $10x$ the scanning periodicity of the fiber.

\begin{figure}
  \begin{center}
  \includegraphics[width=16cm]{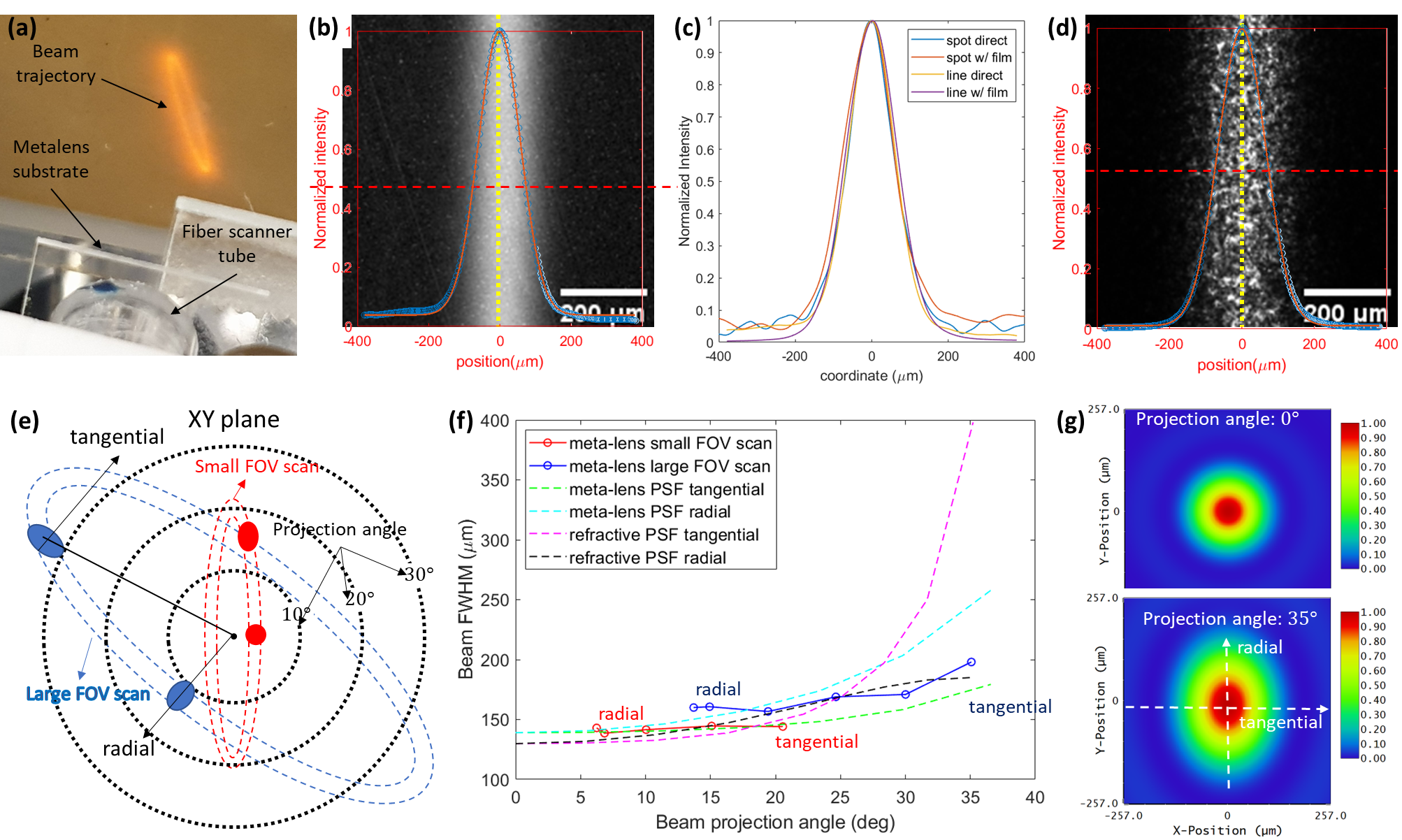}\\
 \caption{\textbf{(a)} Image of the trajectory of the beam projected by the meta-lens onto an IR card. The fiber is driven by a piezo tube at the fiber's mechanical resonance frequency. \textbf{(b,d)} The center part ($6.8 ^{\circ}$ beam projection angle) of the beam trajectory with fiber actuation, captured by the IR microscope at the distance $z=15mm$, (b) is a direct microscope image, (d) is taken with a thin frost film at the focal plane of the microscope (between the microscope and the meta-lens). The intensity distribution across the beam trajectory and averaging along trajectory (the yellow dot line) is measured and plotted. Gaussian blur (radius = $12\mu m$ for direct image and $24 \mu m$ for image with the paper) is applied to reduce the noise. \textbf{(c)} The normalized intensity distributions across the beam spot (without fiber actuation) and the line (the center part of the scanning trajectory, $6.8 ^{\circ}$ beam projection angle), which are measured from the images taken at $z=15mm$ with and without the frost film.\textbf{(e)} Demonstration of the beam scanning trajectory on the illumination plane. \textbf{(f)} The experimentally measured and simulated FWHM of the projected beam vs the projection angle at $z=15mm$. At lower (higher) beam project angle we measure radial (tangential) FWHM. The simulated PSF is calculated using Huygens method in Zemax. \textbf{(g)} PSF of the meta-lens calculated by Huygens method in Zemax for beam projection angle $0^{\circ},35^{\circ}$ at $z=15mm$.
 \label{Fig_scan_traj_measure}} 
  \end{center}
\end{figure}

We actuate the fiber scanner at a resonance frequency of $2187Hz$ with a stable sinusoidal input signal to scan along an elliptical trajectory. This causes the projected beam to traverse along an elliptical trajectory, as can be seen on an IR card (Fig. \ref{Fig_scan_traj_measure}a). Fig. \ref{Fig_scan_traj_measure}b shows the center part of the scanning trajectory ($6.8^{\circ}$ projection angle) taken at $z=15mm$ with an exposure time $>8ms$ . We measure the intensity distribution across the trajectory (averaging along the trajectory line), which is identical to the 1D intensity distribution of the scanning beam at specific positions. As shown in  Fig. \ref{Fig_scan_traj_measure}c, the 1D intensity distribution of the beam spot without fiber actuation (correspond to $0^{\circ}$ projection angle) has no observable difference compared to that of the trajectory line of the scanning beam with a small ($6.8^{\circ}$) projection angle. As the objective lens of our microscope has a small NA of only $0.24$, it cannot directly image the beam with projection angles $>14^{\circ}$. To image the trajectory of the scanning beam at $>14^{\circ}$, a frost film is placed at the focal plane of the microscope to scatter the light. Fig. \ref{Fig_scan_traj_measure}d shows the image of beam trajectory on a frost film. The scattering of the frost film causes a slight broadening ($\sim 15 \mu m$) of the FWHM of the intensity distribution. This excess broadening is removed to recover the actual FWHM.
 
The scanning trajectory at different positions in the x-y plane has different corresponding beam projection angles, as demonstrated in Fig. \ref{Fig_scan_traj_measure}e. 
Therefore, to measure the FWHM of the beam at different projection angles, we move the microscope in the x-y plane to take images of different parts of the elliptical scanning trajectory. We scan over a small and large FOV, whose trajectories are high aspect ratio ellipses that cover the projection angle range $6.2^{\circ} - 20.6^{\circ}$ and $13.7^{\circ} - 35.0^{\circ}$ (Fig.~\ref{Fig_scan_traj_measure}e), respectively. We compared the measured FWHM of the scanning beam vs beam projection angle with the simulation results obtained in Zemax (FWHM of the 1D cross-section of the PSF calculated by Huygens method) in Fig.~\ref{Fig_scan_traj_measure}f. It is worthwhile to notice that at the projection angle $>0^{\circ}$, the beam intensity distribution is not centro-symmetric, due to the oblique incidence of the beam, which results in beam distortions shown in Fig.~\ref{Fig_scan_traj_measure}g. Specifically, two different distributions along two orthogonal axes arise, which we define as radial and tangential axis. The radial axis connects the beam spot to center (intersection of the optical axis and the x-y plane), and the axis perpendicular to that is defined as the tangential axis. As can be seen in Fig. \ref{Fig_scan_traj_measure}e, the intensity distribution across the trajectory is the radial distribution of the scanning beam at the co-vertex of the ellipse (smallest beam projection angle), and is close to the tangential distribution of the beam at near the vertex of the ellipse (large beam projection angle). As can be seen in Fig. \ref{Fig_scan_traj_measure}f, for both small and large FOV scan, the measured FWHMs at the smallest projection angles of the scans (co-vertex position) are close to the simulated radial FWHMs, while at large projection angles of the scans (near to vertex), the measured FWHMs are close to the simulated tangential FWHMs. Therefore, we demonstrate that the measured FWHMs of the scanning beam at various projection angles within the $35^{\circ}$ half-FOV match well with the simulation. According to the simulation, for our meta-lens, at the center of the field ($0^{\circ}$ projection angle), the FWHM of the beam is $\sim 140\mu m$; the tangential and radial FWHM of the beam increase by $29\%$ and $86\%$ respectively at the edge of the FOV ($35^{\circ}$ projection angle). In contrast, when using the state-of-the-art spherical refractive lens doublet, the FWHM of the beam is slightly smaller at the center of the field ($\sim 130\mu m$), but the tangential FWHM of the beam increase by over $187\%$ to $400\mu m$ at the edge of the FOV due to the spherical aberration. The FWHM of the scanning projected beam determines the spatial resolution of our SFE system. Therefore, our meta-lens SFE system can achieve a comparable resolution of the state-of-the-art SFE at the center of the field, but outperforms the state-of-the-art at the edge of the FOV as it suffers significantly less from resolution degradation (less than a factor of 2).

\section{Conclusion}
We realized a meta-lens, suitable for integration and miniaturization of an SFE. We demonstrated an SFE system at $1310nm$ with large FOV and DOF. Our meta-lens SFE system can reach a FOV of $70^{\circ}$, a DOF of $\sim 15mm$, and a spatial resolution of $\sim 140 \mu m$ at  illumination distance up to $15mm$ at the center of the field, which are comparable to the state-of-art. The length of the optics is reduced by $28\%$ (from $1.2mm$ to $0.86mm$), and resolution degradation at the edge of the FOV is reduced to less than a factor of 2. Further device length reduction is possible by using a thinner substrate and increasing the NA of the fiber. We also note that, while our current implementation employs a single wavelength, using polychromatic meta-lenses operating in red, green and blue wavelengths, full color visible SFE can be realized.

\section*{Disclosures}
 Eric Seibel is the consultant to VerAvant Inc., Redmond, WA, which has University of Washington license for the SFE

\section*{Acknowledgements}
The work is supported by NSF GCR 2120774. Part of this work was conducted at the Washington Nanofabrication Facility / Molecular Analysis Facility, a National Nanotechnology Coordinated Infrastructure (NNCI) site at the University of Washington with partial support from the National Science Foundation (NSF) via awards NNCI-1542101 and NNCI-2025489. We thank Dr. Luis Savastano for providing SFE angioscope image in Figure 1.

\bibliographystyle{abbrvnat}
\bibliography{references}  %%% Uncomment this line and comment out the ``thebibliography'' section below to use the external .bib file (using bibtex) .





% \end{thebibliography}

\end{document}